\documentclass{aa}  

\usepackage[utf8]{inputenc}
\usepackage[T1]{fontenc}
\usepackage{multirow}
\usepackage{xcolor}
\usepackage{natbib}
\usepackage{mathtools}
\usepackage{hyperref}
\usepackage{graphicx}
\usepackage{dcolumn}
\usepackage{bm}
\usepackage{siunitx}	
\usepackage{physics}
\usepackage{comment}
\DeclareSIUnit \parsec {pc}

\usepackage{txfonts}
\usepackage[switch]{lineno} 
\usepackage[normalem]{ulem}
\usepackage{placeins}

\begin{document} 

   \title{Examining extinction distributions for type Ia supernovae in simulated 3D galaxies}

\author{
J.~Duarte \inst{1,}\thanks{joao.r.d.duarte@tecnico.ulisboa.pt},
S.~Gonz\'alez-Gait\'an \inst{2},
A.~M.~Mour\~ao \inst{1},
R.~P.~Santos \inst{1,3},
R.~Wojtak \inst{4}
}

\authorrunning{J.~Duarte}
\institute{CENTRA- Centro de Astrofísica e Gravitação, Instituto Superior T\'ecnico, Universidade de Lisboa, Av. Rovisco Pais 1, 1049-001 Lisboa, Portugal
\and
Instituto de Astrof\'isica e Ci\^encias do Espaço, Faculdade de Ci\^encias, Universidade de Lisboa, Ed. C8, Campo Grande, 1749-016 Lisbon, Portugal
\and
European Southern Observatory, Alonso de C\'ordova 3107, Casilla 19, Santiago, Chile
\and
DARK, Niels Bohr Institute, University of Copenhagen, Jagtvej 155, 2200 Copenhagen, Denmark
}
   \date{Accepted July 2026/ Received May 2026
}
 
\abstract
{Dust extinction and reddening, commonly parameterized by $A_V$ and $E(B-V)$, respectively, greatly contribute to type Ia supernovae (SNe Ia) observed color and magnitude variations. The models used to describe the extinction probability density function (PDF) are often simplistic and outdated, which can negatively impact SN simulations and cosmology, adding to systematic errors.}
{We present an analysis of simulated SN Ia extinction in galaxies along realistic lines of sight and investigate the parameterization of its PDF $p(A_V)$, as well as its dependence on host properties, such as morphology and dust mass.}
{We employed SKIRT, a radiative transfer code, to simulate observations of SNe Ia in various locations of different galaxies and generate synthetic extinction distributions. To parameterize and fit these distributions, we used both the commonly assumed single-parameter exponential PDF and some of its two-parameter generalizations.}
{We find that the standard exponential PDF does not adequately describe simulated SN extinction: It underestimates low-extinction events and overestimates high-extinction ones. Furthermore, 2D Kolmogorov-Smirnov tests show significant differences between the simulated extinction distributions for SNe in different environments (e.g., bulges or spiral arms), which the exponential parameterization cannot properly distinguish. In contrast, the two-parameter PDFs parameterize SN extinction distributions more accurately across all simulated environments. Variations in host morphology (i.e., SN and dust distributions) or dust mass relate to variations in different PDF parameters, meaning that the two-parameters descriptions offer improved sensibility to host differences.}
{We conclude that the two-parameter Weibull or exponentiated exponential PDFs offer the best overall parameterizations of SN Ia extinction for a wide range of simulated environments. Analyzing observed SN color data from the literature and assuming a Gaussian distribution for the intrinsic component, we conclude that a two-parameter extinction PDF results in SNe being intrinsically redder, with their mean intrinsic color shifted $\sim2\sigma$ in relation to the standard exponential extinction PDF.}


\maketitle
\section{Introduction}
\label{sec:intro}

Type Ia supernovae (SNe Ia) are bright transient events that occur in all galaxy types. Their standardizable luminosity makes them ideal distance indicators for cosmology, which made them instrumental in the discovery of the accelerated expansion of the Universe \citep{Riess98,Perlmutter_99}. They are frequently used to constrain cosmological parameters, such as the Hubble constant $H_0$ \citep[e.g.,][]{Riess_2022}. 

\par
As with most extragalactic sources, SN Ia light interacts with environmental dust as it travels across its host galaxy, undergoing both absorption and scattering. Due to these effects, a SN will appear both fainter and redder in proportion to the amount of dust along its line of sight, which is usually quantified by the dust extinction $A_V$. The reddening effect, in particular, is a consequence of the light/dust interaction cross-sections, which are higher for bluer wavelengths and depend on the size and composition of the dust particles. As such, the observed SN color $c$ is actually a combination of the intrinsic color $c_{\mathrm{int}}$ and the extinction-induced color excess $E(B-V)$, as described in Eq. \ref{eq:color}:

\begin{equation}
    c \ = \ c_{\mathrm{int}} \ + \ E(B-V).
    \label{eq:color}
\end{equation}
\par
The shifts caused by dust extinction on SN color and magnitude mean that it must be accounted for in any study dealing with SNe Ia. Given that it can be difficult to directly gauge the extinction for individual SNe, knowing the correct shape of the extinction probability density function (PDF) is essential, particularly when working with large observational samples or simulations.

\par

In cosmology, for example, SNe Ia are standardized by taking advantage of two well-known empirical light-curve relations: the shape-luminosity relation \citep{Phillips_1993} and the color-luminosity relation \citep{Tripp98}, with dust playing an important role in the latter. For a given SN with observed peak magnitude $m_B$, stretch $x_1$, and peak color $c$, the standardized magnitude $m_B^{corr}$ can be calculated from Eq. \ref{eq:tripp}:

\begin{equation}
    \label{eq:tripp}
    m_B^{corr} \ = \ m_B + \alpha  x_1 - \beta  c + \delta_M,
\end{equation}
\noindent
where $\alpha$ and $\beta$ describe the linear relation of brightness with stretch and color, respectively, and $\delta_M$ is the ``mass step,'' an ad hoc term introduced to account for the fact that, otherwise, corrected luminosities for SNe Ia originating in high-mass galaxies would be systematically higher than for those in low-mass hosts \citep{Sullivan10, Lampeitl_2010, kelly_2010}.
\par

It is possible to define a more complete color model by separating the observed color into its components according to Eq. \ref{eq:color}, so that the color correction is expressed in terms of the intrinsic color-luminosity relation $\beta_{\mathrm{int}}$ and the environment-dependent total-to-selective extinction ratio $R_V$, as shown in Eq. \ref{eq:color_corr}:

\begin{equation}
    \beta c \ \rightarrow \ \beta_{\mathrm{int}} \ c_{\mathrm{int}} + (1+R_V) \ E(B-V).
    \label{eq:color_corr}
\end{equation}

\noindent
Defining the color correction in this way highlights a fundamental problem with the standard $\beta c$ correction, as it shows that the relation between observed color and observed magnitude is not linear and therefore not well represented by a single $\beta$ value.

\par
When using this more complete color model in a Bayesian framework, the extinction PDF is often employed as a prior to inform the overall shape of the fitted posterior distributions, both for individual light-curve fits \citep[e.g.,][]{Mandel_2022} and for large-sample cosmological fits \citep[e.g.,][]{Popovic_2023}. Different choices for the extinction PDF can have wide-ranging effects, not only on $E(B-V)$ and $c_{\mathrm{int}}$, but also on the subsequent cosmological results, given that $c_{\mathrm{int}}$ and $E(B-V)$ have different contributions to the color-luminosity relation. In general, $\beta_{\mathrm{int}} < 1+R_V$ \citep{Mandel_2017,Brout_2021,gaitan_2021}, meaning that, for two SNe with the same observed color, the one with higher reddening will appear more dimmed.

\par
The extinction PDF also plays an important role in SN Ia simulations. In bias correction simulations \citep[e.g.,][]{Jha_2007,Kessler_2009,Kessler_2017,Popovic_2021,Vincenzi_2021}, which are used to quantify and correct for a variety of effects, ranging from Malmquist \citep[e.g.,][]{Marriner_2011} and SN Ia light-curve selection biases \citep[e.g.,][]{Kessler_2017} to effects arising from empirical relations between SNe and their host galaxies \citep[e.g.,][]{Popovic_2021}, the relative contribution of dust extinction is crucial to determine whether a SN with a given color sits above the detection threshold. Similarly, in simulations of SN rates \citep[e.g.,][]{Perrett_2012,Rodney_2014,wiseman_2021}, the number of events recovered by the simulated detectors can shift depending on the amount of light obscured by dust.

\par
It is generally assumed that dust extinction follows an exponential PDF \citep[e.g.,][]{Jha_2007,Kessler_2009a,Holwerda_2014,Mandel_2017} or, less commonly, the sum of an exponential and Gaussian PDFs \citep[e.g.,][]{Neill_2006,Rodney_2014}. However, these prescriptions are based on limited and, in some cases, outdated simulations \citep[e.g.,][]{Hatano_1998,COMMINS_2004,Riello_2005} and it has been suggested that they do not accurately describe the shape of the true $E(B-V)$ distribution \citep[e.g.,][]{ward_2023,wojtak_2023,hallgren2025}. In addition, a universal extinction PDF is typically assumed for all SNe Ia, neglecting possible environmental differences between different SN populations, which many authors have pointed out \citep[e.g.,][]{gaitan_2021, Brout_2021, wiseman_2022,Duarte_2023, popovic2024modelling,wojtak_2025,Martins_2025}. As such, a more in-depth investigation into the nature of SNe Ia extinction inside host galaxies is required.

\par
In this work, we provide a detailed analysis of simulated SNe Ia dust environments, with the goal of improving upon the description of $A_V$ and $E(B-V)$ distributions\footnote{Given that we will be working with a single dust type, these distributions are equivalent up to a constant $R_V = 3.068$ \citep{Duarte_2025}.}. We simulate SNe Ia embedded in galaxies and infer their respective dust extinction. To these data, we fit several analytical PDFs, with the aim of determining which of them best describes the simulated extinction distributions. The expressions for these PDFs are detailed in Section \ref{sec:methods}, along with the SNe Ia simulation framework. This framework is based on the radiative transfer code SKIRT \citep{Camps_2015,CAMPS_2020}, which is used to simulate the interaction of SN Ia light with dust along various lines of sight. We analyze the extinction of SN populations occurring in different environments, such as elliptical and spiral galaxies, differentiating, in the latter case, between SNe occurring in the disk and the bulge. In Section \ref{sec:results}, we compare the various fitted PDFs and comment on whether they are suitable to describe extinction. We also study how factors such as environment morphology and dust mass affect the simulated extinction distributions for the SNe. Finally, we explore the effects of assuming different $E(B-V)$ parameterizations on the determination of SNe Ia intrinsic color, based on real observed SN color distributions. In Section \ref{sec:discussion} we discuss the impacts that a particular extinction PDF has on SN Ia research, whether it be on bias correction or SN rate simulations or on light-curve and cosmological fits. We also compare our simulations with literature results for SN extinction distributions. Finally, in Section \ref{sec:conclusions} we present our main conclusions.

\section{Methods}
\label{sec:methods}
\subsection{SN Ia simulation}
\label{sec:gal_sim}
In this work, we used radiative transfer models to simulate observations of SNe Ia inside different galactic environments. We used the Monte Carlo-based SKIRT code \citep{Camps_2015,CAMPS_2020}\footnote{\url{https://skirt.ugent.be/root/_home.html}}. By accounting for both scattering and absorption effects, SKIRT can reliably trace the path of photons through a dusty medium to simulate observations for any given sight line.
\par
We simulated observations for three distinct SNe Ia samples, each corresponding to a different environment: SNe occurring in an elliptical galaxy, SNe occurring in a spiral disk, and SNe occurring in a spiral bulge. We constructed a model of the dust distribution inside each host galaxy type and placed the SNe in random positions inside that continuous dust environment, following the host stellar density distribution, which has been shown to trace SN Ia abundances \citep{Anderson_2015,Pritchet_2024}. For simplicity, we ignored dust clumpiness and increased localized dust attenuation in star-forming regions. Although these scenarios would surely have a large impact on our results, our aim is to understand the basic geometry.

\par
A total of $500$ SNe were simulated for each of the three populations, and each object was observed along 30 random lines of sight. As such, in practice, our sample consists of 15000 SNe Ia for each of the three environments. This number was chosen to optimize the computational time of the simulations while, at the same time, ensuring an adequate sampling of the extinction PDFs. For each SN line of sight, SKIRT's ``OligoExtinctionOnly'' mode was used to simulate the path of $10^6$ photon packets with wavelengths of $\lambda=0.445\si{\micro\meter}$ ($B-$band) and $\lambda=0.551\si{\micro\meter}$ ($V-$band). By directly comparing the original emitted flux with the radiative transfer output, we computed unambiguous values for $A_V$ and $E(B-V)$. Each SN was observed from a simulated detector located at a distance of $10\si{\mega\parsec}$ and with an aperture of $1$ AU. We chose such a small aperture to mitigate, as much as possible, the scattering of light into the line of sight. This type of scattering is not very relevant in actual SNe Ia observations, as their short lifespans limit the radius inside which light can effectively scatter back into the line of sight.

\par
The host galaxy models used were based on those described in \cite{Duarte_2025}, summarized here with some minor modifications. These are low-dimensional idealized models meant to describe the overall structure of SN Ia host galaxies. A \cite{Bruzual_2003} spectral energy distribution (SED) was assumed for the SNe Ia, taking a \cite{chabrier_2003} initial mass function (IMF), with a metallicity $Z=0.02$ and an age of $t_{age}=5$Gyr. While this description is not representative of actual SNe Ia, this is not important for the present study, as the recovered point-source extinction does not depend on the point-source SED (see \cite{Duarte_2025}), only on the spatial distribution of the objects and the dust. As stated above, both the original and dust-extincted SEDs are known, which allows for an unambiguous determination of the extinction law. For the dust medium, a \cite{Zubko_2004} dust model was used, corresponding to a total-to-selective extinction ratio $R_V=3.068$ \citep{Duarte_2025}. The morphology of the host galaxies is detailed in the following sections, with their structural parameters summarized in Tab. \ref{tab:morph}.
\par
While this paper focuses on the impacts of dust on SNe Ia, the extinction PDFs we recovered are also valid for other point sources, such as white dwarfs, as long as they follow a spatial distribution similar to the one assumed here. Our approach can also be generalized to other objects, provided that the correct event density distributions are specified.

\par

\def\arraystretch{1.5}
\begin{table}[]
    \centering
    \caption{Structural parameters of the host galaxy stellar and dust distributions for each of the SN environments.}
    \begin{tabular}{c c c}
    \hline\hline
        SN Environment & Parameter & Value \\ \hline
    \multirow{4}{*}{Spiral Disk}  &  $h_{R}^*$  &  $4000\si{\parsec}$\\
      & $h_{z}^*$  &  $350\si{\parsec}$ \\
      & $h_{R}^{\mathrm{D}}$  &  $4000\si{\parsec}$\\
      & $h_{z}^{\mathrm{D}}$  &  $250\si{\parsec}$ \\\hline
     \multirow{5}{*}{Spiral Bulge}     & $r_{\mathrm{eff}}^*$ & $1600\si{\parsec}$ \\
      & $q^*$ & $0.7$ \\
      & $n^*$ & $2$ \\
      & $h_{R}^{\mathrm{D}}$  &  $4000\si{\parsec}$\\
      & $h_{z}^{\mathrm{D}}$  &  $250\si{\parsec}$ \\\hline

      \multirow{5}{*}{Elliptical Galaxy} & $r_{\mathrm{eff}}^*$ & $4000\si{\parsec}$\\
      & $q^*$ & $0.5$ \\
      & $n^*$ & $4$ \\
      & $c^{\mathrm{D}}$ & $4000\si{\parsec}$ \\
      & $q^{\mathrm{D}}$ & $0.5$ \\ \hline

    \end{tabular}

    \label{tab:morph}
\end{table}
\def\arraystretch{1}
\subsubsection{SNe in a spiral disk}
\label{sec:methods_disk}
Positions for the SNe in the spiral disk were sampled from a double-exponential profile, which also describes the spiral galaxy dust distribution. This profile is given by Eq. \ref{eq:exp_disk}:

\begin{equation}
    \rho^{*,\mathrm{D}}(R,z)=\rho^{*,\mathrm{D}}_0\exp(-\frac{R}{h^{*,\mathrm{D}}_R}-\frac{\abs{z}}{h^{*,\mathrm{D}}_z}),
    \label{eq:exp_disk}
\end{equation}
\noindent
where $\rho^*$ and $\rho^{\mathrm{D}}$ refer to the stellar and dust profiles, respectively. The stellar profile was characterized by a scale length $h_{R}^*=4000\si{\parsec}$ and a scale height $h_{z}^*=350\si{\parsec}$, and was normalized by a stellar density $\rho_{0}^*$. These values roughly correspond to those obtained by \cite{De_Geyter_2014} for a sample of observed spiral galaxies. The dust medium was characterized by a dust scale length $h_{R}^{\mathrm{D}}=4000\si{\parsec}$ and a dust scale height $h_{z}^{\mathrm{D}}=250\si{\parsec}$, following once again \cite{De_Geyter_2014}.
\par
The normalization constant $\rho_{0}^{\mathrm{D}}$ was defined via the total dust mass of the system, an input parameter of the simulation that was varied to produce different host environments. From the mean ratio between the scale-length (for either the stellar or dust profiles) and the optical radius $R_{25}$, as reported by \cite{Casasola_2017}, and the scaling relations between $R_{25}$ and the stellar mass, recovered by \cite{Pilyugin_2021}, we estimated the stellar mass of the simulated host galaxy to be $\log(M_{*}/M_{\odot})\sim10$.

\subsubsection{SNe in a spiral bulge}
\label{sec:methods_bulge}

Positions for the SNe in the spiral bulge were drawn from a flattened spheroidal \cite{sersic_63} profile, as defined by Eq. \ref{eq:sersic}:

\begin{equation}
    \rho^*(R,z)=\frac{\rho^*_0}{q^*} \ S_{n^*}\left(\frac{1}{r^*_{\mathrm{eff}}}\sqrt{R^2 + \frac{z^2}{{q^*}^2}}\right),
    \label{eq:sersic}
\end{equation}

\noindent
 with stellar bulge flattening parameter $q^*=0.7$ and stellar bulge effective radius $r_{\mathrm{eff}}^*=1600\si{\parsec}$. $S_n(s)$ is the Sérsic function of order $n$, $n^*=2$ is the stellar bulge Sérsic index \citep{sersic_63}, and $\rho_0^*$ is the stellar bulge density normalization parameter. Once again, these values roughly correspond to those obtained by \cite{De_Geyter_2014} for a sample of observed spiral galaxies.

\par
As described in Section \ref{sec:methods_disk}, the spiral galaxy dust medium was defined by a double-exponential profile. We assumed that this was the only dust interacting with SNe located in the bulge.

\subsubsection{SNe in an elliptical galaxy}
\label{sec:methods_elliptical}

For the elliptical galaxy, the SN positions were selected from a flattened spheroidal \cite{sersic_63} profile, defined by Eq. \ref{eq:sersic}, with flattening parameter $q^*=0.5$, effective radius $r_{\mathrm{eff}}^*=4000\si{\parsec}$, and Sérsic index $n^*=4$. These values roughly follow those cited by \cite{Beifiori_2012} for a sample of observed elliptical galaxies.
\par
The elliptical galaxy dust medium was described by a flattened spheroidal \cite{Plummer_1911} density profile, based on \cite{Baes_2000} and given by Eq. \ref{eq:sphere}:

\begin{equation}
    \rho^{\mathrm{D}}(R,z)=\frac{\rho^{\mathrm{D}}_0}{q^{\mathrm{D}}} \  \left[1 + \frac{1}{{c^{\mathrm{D}}}^2}\left( R^2 + \frac{z^2}{{q^{\mathrm{D}}}^2}\right) \right]^{-5/2}
    \label{eq:sphere}
\end{equation}

\noindent
with the dust flattening parameter $q^{\mathrm{D}}=0.5$, the dust scale length $c^{\mathrm{D}}=4000\si{\parsec}$, and the dust normalization parameter $\rho_{0}^{\mathrm{D}}$, defined in accordance with the total dust mass of the system \citep{Beifiori_2012}. As with the previous host type, we estimated that the simulated galaxy corresponds to a stellar mass of $\log(M_{*}/M_{\odot})\sim10$.

\subsection{Extinction distribution fits}

It is usually assumed that the $A_V$ and $E(B-V)$ distributions for SNe Ia are well described by an exponential (E) distribution \citep[i.e.,][]{Jha_2007,Kessler_2009a,Holwerda_2014,Mandel_2017}. The expression for this PDF is given by Eq. \ref{eq:dist_exp}:

\begin{equation}
    p_{\mathrm{E}}(A_V; \tau)=\frac{1}{\tau} \exp(\frac{-A_V}{\tau}),
    \label{eq:dist_exp}
\end{equation}
\noindent
where $\tau$ controls the width of the distribution and is most commonly associated with the amount of dust in the lines of sight.
\par

In this work, we proposed and explored alternative parameterizations of the SN Ia extinction PDF. We focused on common two-parameter generalizations of the exponential distribution with slightly different shapes and shape parameters: the exponentiated exponential (EE), whose PDF is given by Eq. \ref{eq:dist_ee} for $\alpha>0$ and which reduces to the E PDF for $\alpha=1$; the exponential-logarithmic (EL), whose PDF is given by Eq. \ref{eq:dist_el} for $0<\theta<1$ and which reduces to the E PDF for $\theta \rightarrow 1$; and the Weibull (W), whose PDF is given by Eq. \ref{eq:dist_w} for $\gamma>0$ and which reduces to the E PDF for $\gamma=1$:

\begin{equation}
    p_{\mathrm{EE}}(A_V; \tau, \alpha)=\frac{\alpha}{\tau} \exp(\frac{-A_V}{\tau})  \left(1-\exp(\frac{-A_V}{\tau})\right)^{\alpha-1};
    \label{eq:dist_ee}
\end{equation}

\begin{equation}
    p_{\mathrm{EL}}(A_V; \tau, \theta)=\frac{1}{\tau \ln(\theta)} \ \frac{(1-\theta)  \exp(\frac{-A_V}{\tau})}{(1-\theta)  \exp(\frac{-A_V}{\tau})-1};
    \label{eq:dist_el}
\end{equation}

\begin{equation}
    p_{\mathrm{W}}(A_V; \tau, \gamma)=\frac{\gamma}{\tau} \ A_V^{\gamma-1}  \exp(\frac{-A_V^{\gamma}}{\tau}).
    \label{eq:dist_w}
\end{equation}

\noindent
In addition, a brief analysis of the gamma distribution is presented in Appendix \ref{app:gamma}, while a deeper exploration can be found in \cite{hallgren2025}.

\par

For each SNe Ia sample, we used a Bayesian framework and a Markov-Chain Monte Carlo (MCMC) to fit each of the PDFs to the simulated extinction data. For this, we relied on the SCIPY\footnote{https://scipy.org/} \citep{2020SciPy-NMeth} and EMCEE\footnote{https://emcee.readthedocs.io/en/stable/} \citep{Foreman-Mackey_2013} Python packages. We employed a standard Gaussian likelihood, defined as a product over all simulated SNe. For each fit, the posteriors were inferred from sampling using 50 walkers, each with 250 iterations and a burn-in of 50 iterations. We adopted flat priors on each of the fit parameters.

\section{Results}
\label{sec:results}
In this section, we present an analysis of extinction distributions for SNe Ia originating in different host environments. We also study the effect that host galaxy dust mass and overall morphology have on these distributions. We test whether the simulated extinction distributions can be properly described by the standard exponential PDF and investigate alternative parameterizations based on two-parameter generalizations of the exponential PDF. Finally, we explore how well the new proposed extinction PDFs can reproduce observed SN Ia colors and what implications they have on the determination of SN intrinsic color.

\subsection{Extinction distributions for SNe Ia}
\label{sec:results_I}

\par
\begin{figure*}
	\centering
	\includegraphics[width=1\textwidth]{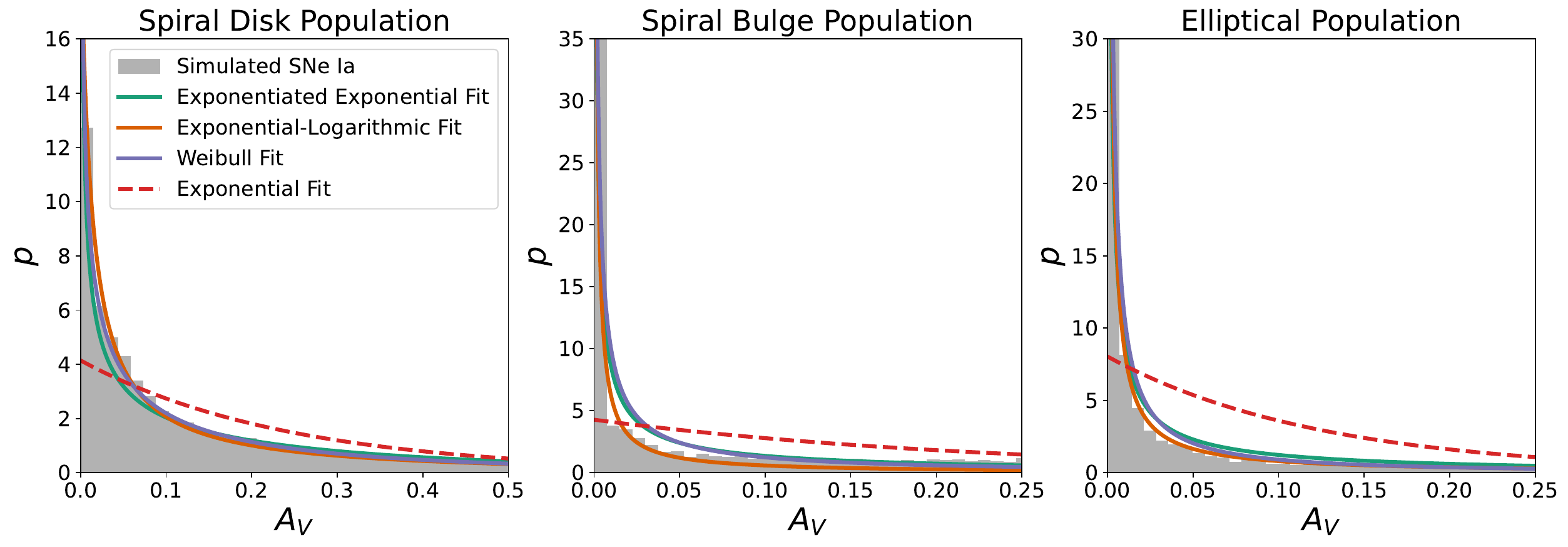}
	\caption{Distributions of $A_V$ for three samples of simulated SNe Ia, occurring in the spiral disk (left), spiral bulge (center), and elliptical galaxy environments (right). The distributions were obtained from radiative transfer simulated observations along random lines of sight for host galaxies with dust mass $M_{\mathrm{D}}=10^7 M_{\odot}$. Best-fit curves for the exponentiated exponential (green), exponential-logarithmic (orange), and Weibull PDFs (purple) are shown, in comparison with the best-fit for the standard exponential PDF (red, dashed).}

\label{fig:av_dist_1e7}
\end{figure*}

We begin by examining the simulated $A_V$ distributions for the three SN Ia samples, located in the spiral disk, the spiral bulge, and the elliptical galaxy. For each host type, we start by assuming a fixed dust mass of $M_{\mathrm{D}}=10^7 M_{\odot}$. The extinction distributions are shown in Fig. \ref{fig:av_dist_1e7}, with best-fit curves for the E, EE, EL, and W PDFs. Corresponding best-fit parameters and Bayesian information criterion (BIC) values are shown in Table \ref{tab:BIC}.
\par
The extinction distributions for the three samples exhibit some common characteristics, the most prominent of which is a large peak for $A_V\sim0$. Even so, their shapes differ quite a lot. To test whether the extinction for the three SN environments follows the same underlying distribution, we performed a two-sample Kolmogorov–Smirnov (KS) test for each pair of normalized extinction distributions. In each case, the $p$-values obtained were consistent with $0$, rejecting the null hypothesis and indicating that the extinction for the three environments cannot be described by the same underlying PDF.

\par
As shown in Fig. \ref{fig:av_dist_1e7} and Table \ref{tab:BIC}, it is evident that the exponential PDF cannot accurately describe any of the simulated distributions, as it favors a much flatter distribution of $A_V$ than is seen in the data. In particular, this parameterization greatly underestimates the prominence of low-extinction SNe, especially in the region of $A_V \sim 0$. At the same time, it predicts an overabundance of high-extinction events, particularly for the SNe in the elliptical galaxy. In addition, even though the SNe in the spiral disk and spiral bulge have significantly different extinction distributions, as indicated by the 2D KS test, the exponential fits return very similar $\tau$ values for both samples. This indicates that the exponential PDF cannot accurately describe the differences between SNe populations from different environments. The fact that the previous results were obtained from idealized galaxy geometries, which represent the best-case scenario for a simple parametric description of SN Ia extinction, further strengthens the point that the exponential PDF is not fit to describe extinction at any level of host structural complexity.

\par
In contrast, the two-parameter PDFs offer much more accurate descriptions of the simulated data, as can be seen in Fig. \ref{fig:av_dist_1e7} and the corresponding BIC values in Table \ref{tab:BIC}. For the SNe in the spiral disk, the best description of dust extinction is given by the Weibull PDF, although it is mostly equivalent to the other two-parameter PDFs. The extinction distributions for SNe in the spiral bulge and the elliptical galaxy are best described by the exponential-logarithmic PDF, which is substantially favored by the BIC. However, for these SNe, the best-fit EL PDFs have values of $\theta\sim 0$ with relatively large uncertainties, which can make it difficult to accurately constrain this parameter. For this reason, using the EE or W distributions might still be best to describe SNe occurring in the spiral bulge or an elliptical galaxy. In all cases, the limits for which each of the two-parameter PDFs reduce to the exponential distribution are strongly ruled out. We add that the gamma distribution, another common two-parameter generalization of the exponential PDF, behaves almost identically to the EE PDF, as we show in Appendix \ref{app:gamma}.

\par

\renewcommand{\arraystretch}{1.5}
\begin{table*}[]
\caption{Best-fit parameters and corresponding BIC values for the different extinction PDFs.}
    \centering
    \begin{tabular}{c c c c c c c}
    \hline\hline
         SN Environment & Model &  BIC & $\tau$ & $\alpha$ & $\theta$ &$\gamma$ \\ \hline 
        \multirow{4}{*}{Spiral Disk} & E & $2.53\times10^{5}$ & $0.241^{+0.002}_{-0.002}$& - & - & - \\
        & EE  & $-1.92\times10^{4}$ & $0.411^{+0.006}_{-0.006}$ & $0.484^{+0.005}_{-0.004}$ & - & -\\
        & EL & $-1.98\times10^{4}$ & $0.625^{+0.012}_{-0.015}$ & - & $0.020^{+0.001}_{-0.001}$ & -\\
        & W & $\bf{-2.04\times10^{4}}$ & $0.163^{+0.002}_{-0.002}$ & - & -&$0.622^{+0.004}_{-0.004}$ \\ \hline
                \multirow{4}{*}{Spiral Bulge} & E & $1.70\times10^{9}$ &  $0.234^{+0.002}_{-0.002}$ & - & - & - \\
        & EE & $1.01\times10^{8}$ & $0.720^{+0.014}_{-0.014}$ & $0.235^{+0.002}_{-0.002}$ & - & - \\
        & EL & $\bf{-5.71\times10^{4}}$ & $0.787^{+0.172}_{-0.274}$  & - & $\left(7.6^{+12200}_{-7.5}\right)\times10^{-8}$ & -\\
        & W & $1.01\times10^{8}$ & $0.090^{+0.002}_{-0.002}$ & - & - & $0.337^{+0.002}_{-0.002}$ \\ \hline
        
        \multirow{4}{*}{Elliptical Galaxy} & E & $9.11\times10^{8}$ & $0.124^{+0.001}_{-0.001}$ & - & - & - \\
        & EE & $-9.54\times10^{4}$ & $0.470^{+0.009}_{-0.009}$ & $0.186^{+0.002}_{-0.002}$ & - & -  \\
        & EL & $\bf{-9.97\times10^{4}}$ & $0.796^{+0.060}_{-0.505}$  & - & $\left(8^{+416}_{-1}\right)\times10^{-6}$ & - \\
        & W & $-9.73\times10^{4}$ & $0.016^{+0.001}_{-0.085}$  & - & - & $0.296^{+0.003}_{-0.003}$ \\ \hline
    \end{tabular}
    \tablefoot{Results for the exponential (E), exponentiated exponential (EE), exponential-logarithmic (EL), and Weibull (W) PDFs are shown. The PDFs were fitted to simulated extinction distributions of SNe Ia along random lines of sight, for different host galaxy types with dust mass $M_{\mathrm{D}}=10^7 M_{\odot}$. Errors on the parameters correspond to a 68\% credible region. For each sample, the lowest BIC value is shown in bold.}
    \label{tab:BIC}
\end{table*}

\subsection{SN Ia extinction distributions and host galaxy dust mass}
\label{sec:dust_mass}
In the previous section, we looked at the extinction distributions for SNe Ia populations in different host galaxies with fixed dust mass and structural parameters. We also explored different parameterizations of these distributions. In this section, we explore the physical meaning of each of the PDF parameters and the way they relate to the host properties, particularly the dust mass.
\par

\begin{figure*}
	\centering
	\includegraphics[width=1\textwidth]{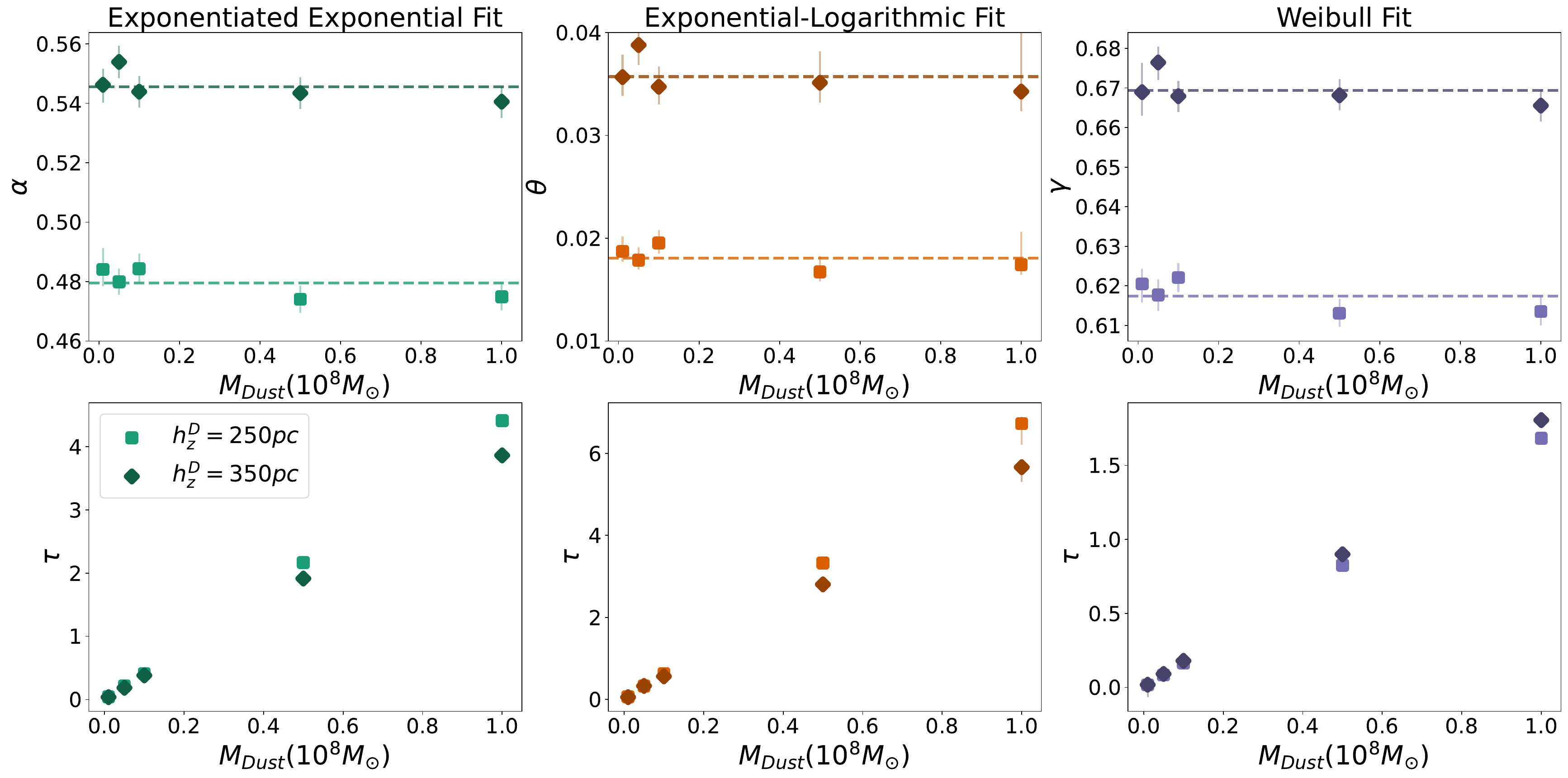}
	\caption{Best-fit parameters for the EE (left), EL (center), and W (right) PDFs as a function of host dust mass $M_{\mathrm{D}}$, with error bars defined by a 68\% credible region. Results for two samples of SNe in spiral disks are shown, corresponding to host galaxies with dust disks of thickness $h_z^{\mathrm{D}}=250\si{\parsec}$ (lighter squares) and $h_z^{\mathrm{D}}=350\si{\parsec}$ (darker diamonds). The dashed lines represent the median $\alpha$, $\theta$ or $\gamma$ values for each fitted PDF and sample.}

\label{fig:av_dist_masses}
\end{figure*}

As an example, we begin by analyzing the impact of dust mass on the extinction distributions for SNe hosted in spiral disks. In Fig. \ref{fig:av_dist_masses}, we plot the best-fit parameters for the EE, EL, and W PDFs for SNe in galaxies with different discrete dust masses $M_{\mathrm{D}}$. We note that, while only the spiral disk is discussed, the observed behavior is consistent across the three simulated samples.

\par
For the three fitted PDFs, $\tau$ is shown to increase linearly with $M_{\mathrm{D}}$, analogously to what is expected for the exponential PDF. Given that $A_V$ is proportional to the amount of dust along the line of sight, for samples with fixed SN and dust distributions, varying the dust mass of the host results in PDFs with the same shapes but different scales, as parameterized by $\tau$. In contrast, while there are some fluctuations in $\alpha$, $\theta$, and $\gamma$, we find that these parameters remain fairly constant across the entire range of host dust masses. The small variations that do exist are consistent across the different fitted PDFs and most likely stem from stochastic variations in the simulation data. We note that these parameters appear to relate in some way to overall host morphology and their role will be further explored in Section \ref{sec:morph}.

\par
Having established a baseline for the behavior of the extinction distributions, we now extend our analysis to a more realistic scenario, given that observed SN samples are not typically made up of events occurring in a single host with a fixed dust mass, but rather in galaxies with differing amounts of dust. As such, we include dust mass as a random parameter in our simulations, drawn from the host mass distributions available from DustPedia\footnote{DustPedia is an online archive of model derived physical parameters for a large sample of galaxies observed with multiwavelength imagery, which can be accessed at \url{http://dustpedia.astro.noa.gr/}.}, which are plotted in Fig. \ref{fig:dust_mass_dist}. For spiral galaxies, DustPedia dust masses can be as low as $M_{\mathrm{D}}=10^4M_{\odot}$, but mostly range from $M_{\mathrm{D}}=10^6M_{\odot}$ to $M_{\mathrm{D}}=10^9M_{\odot}$. Assuming, according to the scaling relations discussed in Section \ref{sec:methods_disk}, an estimated stellar mass of $M_*\sim 10^{10}$ for the SKIRT galaxy models, the dust-to-stellar mass ratios $\log(M_{\mathrm{D}}/M_*)$ for the simulated SN spiral hosts will be between $-4$ and $-1$. These values are in line with those observed by \cite{Smith_2012} and \cite{De_Geyter_2014}. For elliptical galaxies, the DustPedia dust masses range from $M_{\mathrm{D}}=10^4M_{\odot}$ to $M_{\mathrm{D}}=10^8M_{\odot}$. As such, the dust-to-stellar mass ratios for the simulated SN elliptical host galaxies will be between $-6$ and $-2$. In general, this range agrees with the values observed by \cite{Smith_2012}. It also includes some dustier galaxies, which are not very prevalent and are in line with the upper end of observed elliptical dust-to-stellar ratios \citep{Rowlands_2012,Michalowski_2019,Lesniewska_2023}.

\par
\begin{figure*}
	\centering
	\includegraphics[width=1\textwidth]{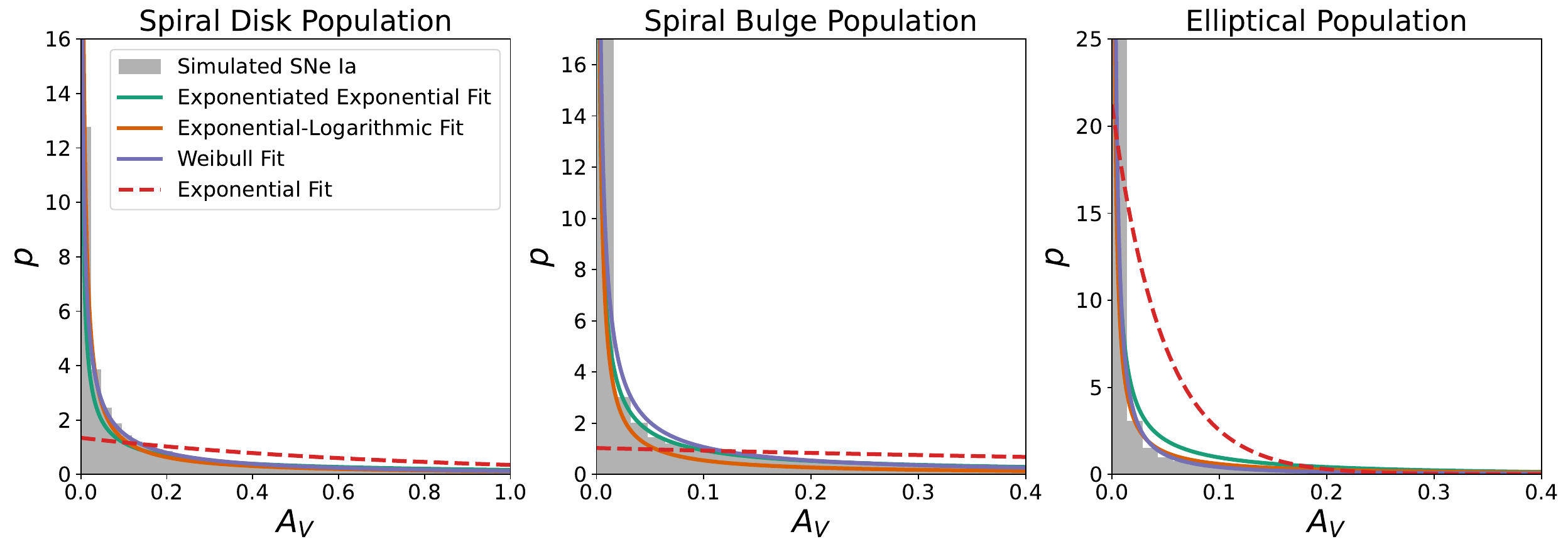}
	\caption{Same as Fig. \ref{fig:av_dist_1e7}, but for SNe hosted in galaxies with random dust masses drawn from DustPedia distributions, ranging from $M_{\mathrm{D}}=10^4M_{\odot}$ to $M_{\mathrm{D}}=10^9M_{\odot}$.}

\label{fig:av_dist_all_mass}
\end{figure*}

The SN extinction distributions obtained under these conditions for the three populations are shown in Fig. \ref{fig:av_dist_all_mass}. Best-fit curves for the E, EE, EL, and W PDFs are also plotted. The corresponding best-fit parameters and BIC values for each distribution are shown in Table \ref{tab:BIC_all_mass}. The overall shape of the distributions remains consistent with that of the fixed host mass samples. Similarly, we find that the use of a realistic dust mass distribution does not negatively impact the fitting power of the two-parameter PDFs. The BIC values confirm that, as was the case for a host with $M_{\mathrm{D}}=10^7M_{\odot}$, the W PDF is the best at describing the extinction for SNe in spiral disk, while the extinction for SNe in the spiral bulge and the elliptical galaxy is best described by the EL PDF. In the case of the SNe in the spiral bulge, the EL distribution fit cannot properly converge to a precise $\tau$ value, which happens because the value of $\theta$ is very close to 0.

\par

\begin{table*}[]
\caption{Similar to Tab. \ref{tab:BIC}, but for random SN Ia host galaxy dust masses.}
    \centering
    \begin{tabular}{c c c c c c c}
    \hline\hline
        SN Environment & Model &  BIC & $\tau$ & $\alpha$ & $\theta$ & $\gamma$ \\ \hline
        \multirow{4}{*}{Spiral Disk} & E & $3.02\times10^9$ & $0.744_{-0.002}^{+0.002}$ & - & - & -\\
        & EE  &$-2.95\times10^4$ & $2.07_{-0.02}^{+0.02}$ & $0.282_{-0.001}^{+0.001}$ & - & -
\\
        & EL & $-3.87\times10^4$ & $3.67_{-0.95}^{+0.05}$ & - & $\left(5.0_{-0.1}^{+0.2}\right)\times 10^{-4}$&-\\

        & W &$\bf{-4.56\times10^4}$ & $0.242_{-0.002}^{+0.002}$ & - & - & $0.435_{-0.001}^{+0.001}$\\ \hline
                \multirow{4}{*}{Spiral Bulge} & E & $8.79\times10^9$& $0.969_{-0.004}^{+0.004}$ & - & - & -\\
        & EE & $1.16\times10^8$ & $3.68_{-0.03}^{+0.03}$ & $0.190_{-0.001}^{+0.001}$\\
        & EL & $\bf{-1.25\times10^5}$& $4.6_{-3}^{+5}$ & - &$\left(1.2_{-0.1}^{+138}\right)\times 10^{-8}$ & -\\
        & W & $1.16\times10^8$ & $0.187_{-0.002}^{+0.002}$ & - & - & $0.296_{-0.001}^{+0.001}$\\ \hline
        
        \multirow{4}{*}{Elliptical Galaxy} & E & $3.20\times10^9$ & $\left(4.71_{-0.02}^{+0.02}\right)\times 10^{-2}$ & - & - & - \\
        & EE & $1.17\times10^7$ & $0.265_{-0.003}^{+0.003}$ & $0.139_{-0.001}^{+0.001}$ & - & -\\
        & EL & $\bf{-8.21\times10^5}$& $0.48_{-0.21}^{+0.01}$ & - & $\left(3.1_{-0.1}^{+80}\right)\times 10^{-7}$ & -\\
        & W &$1.17\times10^8$ & $\left(1.69_{-0.03}^{+0.03}\right)\times 10^{-3}$ & - & - & $0.258_{-0.001}^{+0.001}$ \\ \hline
    \end{tabular}
    \tablefoot{Host galaxy dust masses were drawn from the DustPedia dust mass distributions, which range from $M_{\mathrm{D}}=10^4M_{\odot}$ to $M_{\mathrm{D}}=10^9M_{\odot}$.}
    \label{tab:BIC_all_mass}
\end{table*}
\par
We note that, while the host galaxy structural parameters stayed consistent between the fixed- and random-dust mass samples, the values of the fitted morphological parameters ($\alpha$, $\theta$ and $\gamma$ in Eqs. \ref{eq:dist_ee}-\ref{eq:dist_w}) varied considerably. This shows that even the mixing of SNe populations from different host galaxies, with the same morphology but different dust masses, can substantially impact the shape of the extinction distributions. 

\subsection{SN Ia extinction distributions and host morphology}
\label{sec:morph}
In the previous section, we showed that variations in the dust mass of SN Ia host galaxies are tied to variations in $\tau$, as seen in Fig. \ref{fig:av_dist_masses}. We also showed that $\alpha$, $\theta$, and $\gamma$ are constant across different dust masses for single-host samples, but can change for samples including multiple hosts with the same morphology and different dust masses. In this section, we analyze the relation between host morphology and the extinction distribution.

\par
As an example, we again focus on the impacts of the structural parameters on the extinction distribution for SNe in the spiral disk. To do this, we introduce a new SN sample, whose host galaxy is a slightly modified version of the one described in Section \ref{sec:methods_disk} with a thicker dust disk with $h_z^{\mathrm{D}}=350\si{\parsec}$. We then repeat the process described in Section \ref{sec:dust_mass}, simulating SNe Ia inside this galaxy for different discrete host dust masses $M_{\mathrm{D}}$. The corresponding best-fit parameters for the EE, EL, and W PDFs for these samples are plotted in Fig. \ref{fig:av_dist_masses}.
\par
Comparing the results for the samples hosted in the thick-disk galaxy with those obtained in Section \ref{sec:dust_mass}, we find that $\alpha$, $\theta$, and $\gamma$ change substantially, despite showing no evolution with dust mass in both cases. We therefore posit that these parameters are primarily tied to overall host morphology. The influence of any single structural parameter on the extinction PDF is hard to track, given that galaxy morphology is complex and arises from a combination of multiple structural parameters. However, from this example we find that, in general, lower values of $\alpha$, $\theta$ or $\gamma$ correspond to steeper peaks with a higher prominence of extinction-free SNe. Thus, hosts with more extended dust environments or more centrally located SNe correspond to higher values for these parameters, as seen in Fig. \ref{fig:av_dist_masses}. We also find that $\tau$ is only slightly affected by the change in disk thickness, with higher values found for the thinner disk. However, we conclude that this parameter remains primarily tied to the dust mass.
\par
Once again, it is not realistic to assume that all SNe occur in galaxies with the exact same structural parameters. For this reason, we extended our simulations to incorporate hosts with different morphologies, with a few caveats. The parameter space for observed host galaxy structure is too large to completely probe, even for our relatively simple geometries. In addition, contrary to host galaxy dust mass, there is no complete distribution that indicates how observed geometries may be distributed.
\par
For these reasons, we opt to sample each parameter from a flat distribution, according to the ranges shown in Tab. \ref{tab:morph_range}, which roughly follow the minimum and maximum values reported by \cite{Beifiori_2012} and \cite{De_Geyter_2014} for observed galaxies. For simplicity, the host dust mass was fixed at $M_{\mathrm{D}}=10^7M_{\odot}$. This process is not necessarily meant to reproduce realistic host galaxy populations, but serves rather as a robustness test for the proposed extinction PDFs.

\par
We fitted the E, EE, EL, and W PDFs to the simulated extinction samples generated in this way. The corresponding best-fit parameters and BIC values are shown in Table \ref{tab:BIC_all_morph}. The fits give good and consistent visual results, which, however, are not always confirmed by the BIC values. In particular, for the SNe in the spiral disk, we find that both the EE and W PDFs have higher BIC values than the E PDF, even though they offer a much better visual description of the data. This can be explained by the existence of many SNe with $A_V=0$, which greatly bias the BIC metric. These non-extincted SNe result from hosts with very thin dust disks in comparison to the stellar distributions, which are over-represented in the simulations due to the way in which the structural parameters were drawn. Nevertheless, these SNe do not appear to have a significant impact on the fits, except in the case of the EL PDF fit to the SNe in the spiral bulge, which cannot converge to a proper value for the $\theta$ parameter.
\par

\begin{figure*}
	\centering
	\includegraphics[width=1\textwidth]{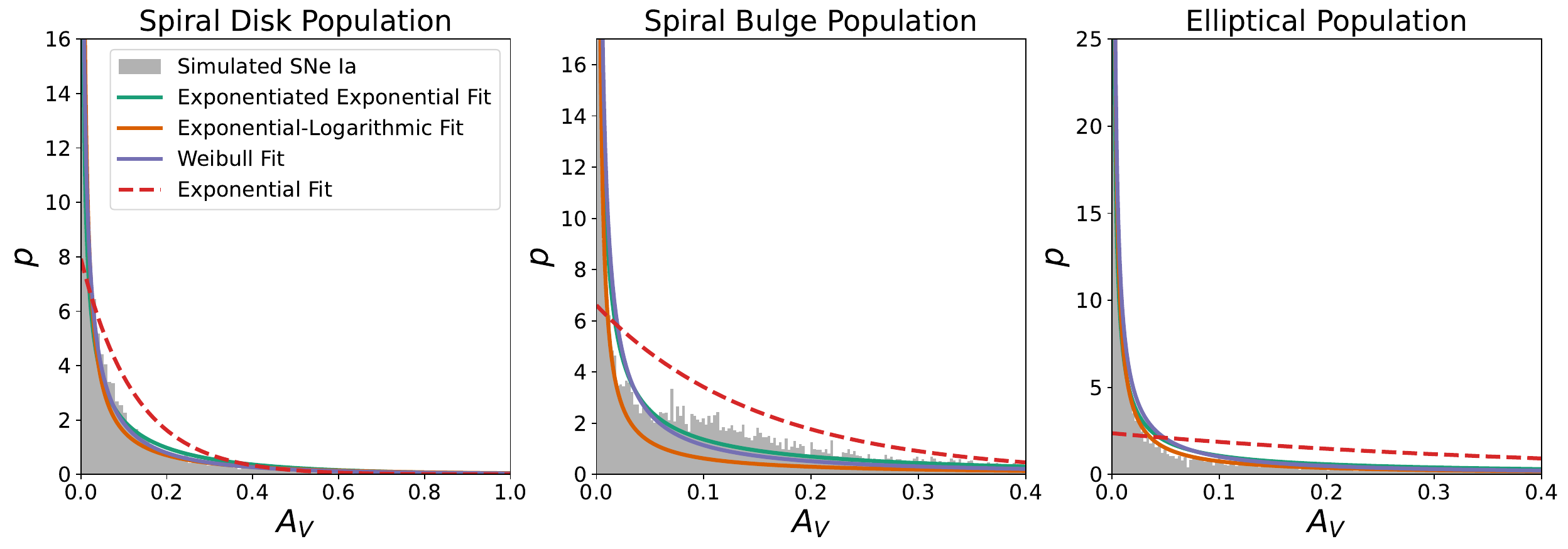}
	\caption{Same as Fig. \ref{fig:av_dist_1e7}, but for host galaxies with random structural parameters, generated according to the parameter ranges recorded in Tab. \ref{tab:morph_range}. Only SNe with $A_V\neq0$ are shown and included in the fits.}

\label{fig:av_dist_all_morph}
\end{figure*}

In an effort to get a fairer assessment of the fit metrics, we repeat our analysis by looking only at SNe with $A_V\neq0$. The respective simulated extinctions are shown in Fig. \ref{fig:av_dist_all_morph}, along with the best-fit distributions. The corresponding best-fit parameters and BIC values are shown in Table \ref{tab:BIC_all_morph}. With the exception of the EL distribution fit for the SNe in the spiral bulge, our sample restriction does not significantly alter the best-fit parameters for any of the distributions. At the same time, this restriction has a great impact on the BIC values, which become more in line with the results discussed in previous sections. We highlight that the exponential PDF is ruled out regardless of whether SNe with $A_V=0$ are included in the sample or not.
\par
We conclude that the two-parameter PDFs are robust and can be used to describe extinction for a diverse sample of SN Ia hosts, even if some unrealistic galaxies are included. We add that a host sample with a more realistic parameter range would represent a subset of the space explored in this section, which the two-parameter PDFs have been shown to easily accommodate.

\def\arraystretch{1.5}
\begin{table}[]
    \caption{Random structural parameter ranges of the host galaxy stellar and dust distributions for each of the SN environments.}
    \centering
    \begin{tabular}{c c c}
    \hline\hline
    SN Environment & Parameter & Range \\ \hline
    \multirow{4}{*}{Spiral Disk}  &  $h_{R}^*$  &  $[3000 \si{\parsec} , 8000 \si{\parsec}]$\\
      & $h_{z}^*$  &  $[100 \si{\parsec}, 1000 \si{\parsec}]$ \\
     & $h_{R}^{\mathrm{D}}$  &  $[3000 \si{\parsec} , 10000 \si{\parsec}]$\\
      & $h_{z}^{\mathrm{D}}$  &  $[100 \si{\parsec}, 500 \si{\parsec}]$ \\\hline
  
    \multirow{5}{*}{Spiral Bulge} & $r_{\mathrm{eff}}^*$ & $[1000 \si{\parsec},6000\si{\parsec}]$ \\
      & $q^*$ & $[0.35,0.9]$ \\
      & $n^*$ & $[1,6]$ \\
      & $h_{R}^{\mathrm{D}}$  &  $[3000 \si{\parsec} , 10000 \si{\parsec}]$\\
      & $h_{z}^{\mathrm{D}}$  &  $[100 \si{\parsec}, 500 \si{\parsec}]$ \\\hline

      \multirow{5}{*}{Elliptical Galaxy} & $r_{\mathrm{eff}}^*$ & $[2000\si{\parsec, 6000\si{\parsec}}]$\\
      & $q^*$ & $[0.4,1]$ \\
      & $n^*$ & $[2,8]$ \\
      & $c^{\mathrm{D}}$ & $[2000 \si{\parsec}, 6000 \si{\parsec}]$ \\
      & $q^{\mathrm{D}}$ & $[0.4, 1]$ \\ \hline

    \end{tabular}
    \label{tab:morph_range}
\end{table}
\def\arraystretch{1}

\renewcommand{\arraystretch}{1.5}
\begin{table*}[]
\caption{Similar to Tab. \ref{tab:BIC}, but for random SN Ia host galaxy structural parameters.}
    \centering
    \begin{tabular}{c c c c c c c}
    \hline\hline
         SN Environment & Model &  BIC & $\tau$ & $\alpha$ & $\theta$ & $\gamma$ \\ \hline
        \multirow{4}{*}{Spiral Disk} & E & $4.64\times10^6$ & $0.126_{-0.001}^{+0.001}$ & - & - & -\\
        & EE  &$5.63\times10^6$ & $0.269_{-0.004}^{+0.004}$ & $0.369_{-0.004}^{+0.004}$ & - & -
\\
        & EL & $\bf{-4.43\times10^4}$ & $0.44_{-0.01}^{+0.01}$ & - & $0.0039_{-0.0002}^{+0.0003}$ &-\\

        & W &$5.63\times10^6$ & $0.068_{-0.002}^{+0.001}$ & - & - & $0.519_{-0.004}^{+0.005}$\\ \hline

                \multirow{4}{*}{Spiral Bulge} & E & $2.03\times10^9$& $0.144_{-0.001}^{+0.001}$ & - & - & -\\
        & EE & $9.86\times10^7$ & $0.485_{-0.009}^{+0.009}$ & $0.218_{-0.002}^{+0.002}$\\
        & EL & $\bf{-8.50\times10^4}$& $1.3_{-0.8}^{+0.4}$ & - & $<1.4\times10^{-6}$ & -\\ 
        & W & $9.86\times10^7$ & $0.045_{-0.003}^{+0.001}$ & - & - & $0.321_{-0.003}^{+0.003}$\\ \hline
    
        \multirow{4}{*}{Elliptical Galaxy} & E & $1.36\times10^9$ & $0.422_{-0.003}^{+0.003}$ & - & - & - \\
        & EE & $-6.07\times10^4$ & $1.55_{-0.03}^{+0.03}$ & $0.184_{-0.002}^{+0.002}$ & - & -\\
        & EL & $\bf{-6.25\times10^4}$& $3.0_{-0.8}^{+0.1}$ & - & $\left(2.2_{-0.2}^{+2}\right)\times 10^{-6}$ & -\\
        & W &$-6.05\times10^4$ & $0.060_{-0.002}^{+0.002}$ & - & - & $0.280_{-0.002}^{+0.002}$ \\ \hline \hline

        \multirow{4}{*}{\shortstack[c]{Spiral Disk \\ $\left(A_V\neq0\right)$}} & E & $4.01\times10^6$ & $0.126_{-0.001}^{+0.001}$ & - & - & -\\
        & EE  &$-4.66\times10^4$ & $0.269_{-0.004}^{+0.004}$ & $0.369_{-0.003}^{+0.003}$ & - & -
\\
        & EL & $-4.40\times10^4$ & $0.44_{-0.01}^{+0.01}$ & - & $0.0039_{-0.0002}^{+0.0003}$ &-\\

        & W &$\bf{-4.72\times10^4}$ & $0.068_{-0.001}^{+0.001}$ & - & - & $0.518_{-0.004}^{+0.004}$\\ \hline

        \multirow{4}{*}{\shortstack[c]{Spiral Bulge \\ $\left(A_V\neq0\right)$}} & E & $1.63\times10^9$& $0.151_{-0.001}^{+0.001}$ & - & - & -\\
        & EE & $\bf{-7.06\times10^4}$ & $0.485_{-0.009}^{+0.009}$ & $0.218_{-0.002}^{+0.002}$\\
        & EL & $-6.13\times10^4 $& $1.27_{-0.5}^{+0.05}$ & - &$\left(2.08_{-0.2}^{+70}\right)\times 10^{-7}$ & -\\
        & W & $-6.77\times10^4$ & $0.045_{-0.002}^{+0.001}$ & - & - & $0.321_{-0.003}^{+0.003}$\\ \hline

        \multirow{4}{*}{\shortstack[c]{Elliptical Galaxy \\ $\left(A_V\neq0\right)$}} & E & $1.36\times10^9$ & $0.423_{-0.003}^{+0.003}$ & - & - & - \\
        & EE & $-6.07\times10^4$ & $1.55_{-0.03}^{+0.03}$ & $0.184_{-0.002}^{+0.002}$ & - & -\\
        & EL & $\bf{-6.25\times10^4}$& $3.0_{-1.4}^{+0.1}$ & - & $\left(2.2_{-0.2}^{+5}\right)\times 10^{-6}$ & -\\
        & W &$-6.05\times10^4$ & $0.060_{-0.003}^{+0.002}$ & - & - & $0.281_{-0.002}^{+0.002}$ \\ \hline
    \end{tabular}
    \tablefoot{Host galaxy structural parameters were generated according to the parameter ranges recorded in Tab. \ref{tab:morph_range}. Results for fits using both the full samples and samples with $A_V\neq0$ are shown.}
    \label{tab:BIC_all_morph}
\end{table*}

\subsection{Impacts of the extinction distribution on the estimation of SN Ia intrinsic color}
\label{sec:pantheon}
In the previous sections, we used radiative transfer simulations to test different parameterizations of the SN Ia extinction distribution. In this section, we explore how well the proposed PDFs comply with SN Ia observations. We also explore how different extinction PDFs affect the estimation of SN intrinsic color $c_{\mathrm{int}}$.
\par
We rely on the Pantheon+SH0ES SNe Ia data set \citep{Scolnic_2022,Brout_2022a,Riess_2022}, combining all duplicate events as described in \cite{wojtak_2025}. The light curves for these SNe were fitted by \cite{brout_2022b} using the SALT2 model \citep{guy_2007,guy_2010} to obtain, among other parameters, the $B-V$ observed color at peak magnitude $c$. As expressed in Eq. \ref{eq:color}, $c$ can be divided into intrinsic and dust components, $c_{\mathrm{int}}$ and $E(B-V)$, respectively. We assume that $c_{\mathrm{int}}$ can be modeled by a Gaussian distribution with mean $\mu_{c_{\mathrm{int}}}$ and standard deviation $\sigma_{c_{\mathrm{int}}}$. For $E(B-V)$, we employ either the E, EE, EL or W PDFs\footnote{In the previous sections, these distributions were used to describe $A_V$, but they are equally valid for $E(B-V)$, given that the two quantities are proportional and $R_V=3.068$ is constant for all the simulated hosts.}. The color distribution is given by a convolution of the $c_{\mathrm{int}}$ and $E(B-V)$ distributions.
\par
\begin{figure*}
	\centering
	\includegraphics[width=1\textwidth]{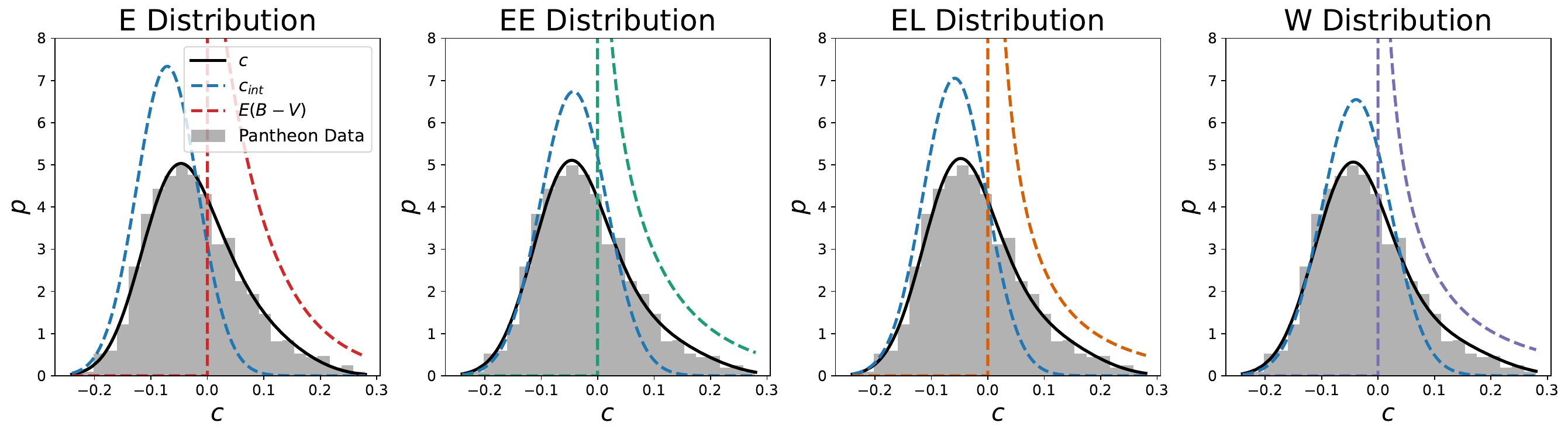}
	\caption{Color distribution for Pantheon+SH0ES SNe Ia. Different best-fit curves to the color distribution are shown in black, assuming a Gaussian distribution for the intrinsic color $c_{\mathrm{int}}$ (blue) and one of the following distributions for the color excess $E(B-V)$: exponential PDF (red, top left), exponentiated PDF (green, top right), exponential-logarithmic PDF (orange, bottom left), or Weibull PDF (purple, bottom right).}

\label{fig:pantheon_fits}
\end{figure*}

The best-fit color distributions for the Pantheon+SH0ES SNe Ia are plotted in Fig. \ref{fig:pantheon_fits}, with the corresponding best-fit parameters and BIC values shown in Tab. \ref{tab:BIC_pantheon}. All the proposed extinction PDFs result in equally good fits of the color data and we find that the proposed two-parameter extinction distributions offer credible reconstructions of the observed color distributions. The fitted values recovered for the morphological parameters $\alpha$, $\theta$, and $\gamma$ are higher than those obtained from the simulations, shown in Tabs. \ref{tab:BIC}, \ref{tab:BIC_all_mass}, and \ref{tab:BIC_all_morph}, suggesting overall more extended dust environments.

\par
Although not very significant ($\sim 2\sigma$), we recover intrinsically bluer SNe when assuming an exponential extinction PDF ($c_{\mathrm{int}}=-0.071_{-0.004}^{+0.004}$) rather than one of the two-parameter generalizations (between $c_{\mathrm{int}}=-0.04_{-0.02}^{+0.02}$ and $c_{\mathrm{int}}=-0.058_{-0.008}^{+0.009}$). As we have shown, the exponential PDF underestimates the prevalence of low-extinction SNe, while overestimating the prevalence of high-extinction events. In the fit, this translates to a bluer intrinsic color. In contrast, when assuming a Gamma PDF for SN Ia extinction \cite{wojtak_2023} recover intrinsically bluer SNe than when an exponential extinction PDF is assumed. The consequences of assuming different extinction PDFs on SN studies and cosmology will be further explored in Section \ref{sec:impacts}.

\begin{table*}[]
\caption{Best-fit parameters and corresponding BIC values for the Pantheon+SH0ES SNe Ia color distribution.}
    \centering
    \begin{tabular}{c c c c c c c c}
    \hline\hline
         Extinction PDF & BIC & $\mu_{c_{\mathrm{int}}}$ & $\sigma_{c_{\mathrm{int}}}$ & $\tau$ & $\alpha$ & $\theta$ & $\gamma$ \\ \hline
        E & $-3264$ & $-0.071_{-0.004}^{+0.004}$ & $0.054_{-0.003}^{+0.003}$ & $0.087_{-0.008}^{+0.008}$ & - & - & - \\
        EE & $-3261$ & $-0.04_{-0.02}^{+0.02}$ & $0.059_{-0.004}^{+0.007}$ & $0.12_{-0.02}^{+0.6}$ & $0.67_{-0.09}^{+0.07}$& -& - \\
        EL & $-3262$& $-0.058_{-0.008}^{+0.009}$ & $0.056_{-0.003}^{+0.003}$ & $0.15_{-0.04}^{+0.1}$ & -& $0.11_{-0.04}^{+0.2}$& - \\
        W & $-3258$ & $-0.04_{-0.02}^{+0.02}$ & $0.061_{-0.006}^{+0.006}$ & $0.09_{-0.02}^{+0.8}$ & -& -& $0.68_{-0.09}^{+0.1}$ \\ \hline

    \end{tabular}
    \tablefoot{A Gaussian distribution, with mean $\mu_{c_{\mathrm{int}}}$ and standard deviation $\sigma_{c_{\mathrm{int}}}$, is assumed for the intrinsic color $c_{\mathrm{int}}$, while the distribution for the color excess $E(B-V)$ is taken to be either an exponential (E), exponentiated exponential (EE), exponential-logarithmic (EL) or Weibull (W) PDF. Errors on the parameters correspond to a 68\% credible region.}
    \label{tab:BIC_pantheon}
\end{table*}

\par
As a consistency test, the above procedure was repeated for some sub-samples of the Pantheon+SH0ES dataset. The first of these was limited to SNe with $z<0.1$, in an effort to minimize selection effects. We find that this restriction does not impact the quality of the fits for any of the parameterizations. In comparison with the full sample fits, we find that the low-redshift PDFs have both a redder mean intrinsic color and larger extinction, as parameterized by $\tau$. We also observe some small deviations in the morphological parameters $\alpha$, $\beta$, and $\gamma$, which are within the error margins described in Tab. \ref{tab:BIC_pantheon}. Overall, these results are in line with what is expected for a low-redshift sample, given the lower level of red SNe missed due to selection biases.
\par
A second subsample, comprising high-stretch SNe with $x_1>0.5$, was also analyzed. Recent studies argue for the existence of multiple populations of SNe Ia, which can have different intrinsic colors \citep[e.g.,][]{gaitan_2021, wojtak_2023,Martins_2025}. Therefore, it is possible that our fitted intrinsic color and extinction distribution parameters are affected by the presence of more than one population in the dataset. The high-stretch sample might represent a cleaner single sub-population \citep[e.g.,][]{wojtak_2023,Martins_2025}, meaning that the red tail of the color PDF can be disentangled and confidently attributed to dust effects \citep[e.g.,][]{wojtak_2025}. For these SNe, the fitted color and extinction PDFs vary little from those obtained from the full sample fit, with the exception of the intrinsic color, which is found to be slightly bluer. This is expected, as the mean observed color for the $x_1>0.5$ subsample is bluer than that observed for the full Pantheon+SH0ES sample.

\section{Discussion}
\label{sec:discussion}

\subsection{Parameterizations of SN Ia extinction distribution}
\label{sec:comparison}
Many attempts have been made to describe and parameterize the extinction distribution of SNe Ia. In this section, we make an overview of some of the most relevant and comment on how they compare to this work.
\par

Despite the prevalence of the exponential parameterization of SN Ia extinction in the literature \citep[e.g.,][]{Jha_2007,Kessler_2009a,Holwerda_2014,Popovic_2021}, there are many studies whose results appear not to support it. Analyzing Monte Carlo simulations of SNe Ia hosted in disk galaxies, \cite{Hatano_1998} find that B-band extinction is very strongly peaked in the region of $A_B<0.1$. Despite favoring an exponential PDF, \cite{Holwerda_2014} report $A_V$ distributions that are not well described by this expression, as it fails to reproduce the strong peak observed for $A_V\sim0$. Likewise, \cite{ward_2023} find that, if no cuts are made to observed SN samples, extinction distributions are more peaked than predicted by the exponential PDF. These results appear to match our own simulations and confirm that the exponential PDF does not offer an adequate description of SN Ia extinction, as it greatly underestimates the probability density of low-extinction events.
\par

\begin{figure}
	\centering
	\includegraphics[width=0.5\textwidth]{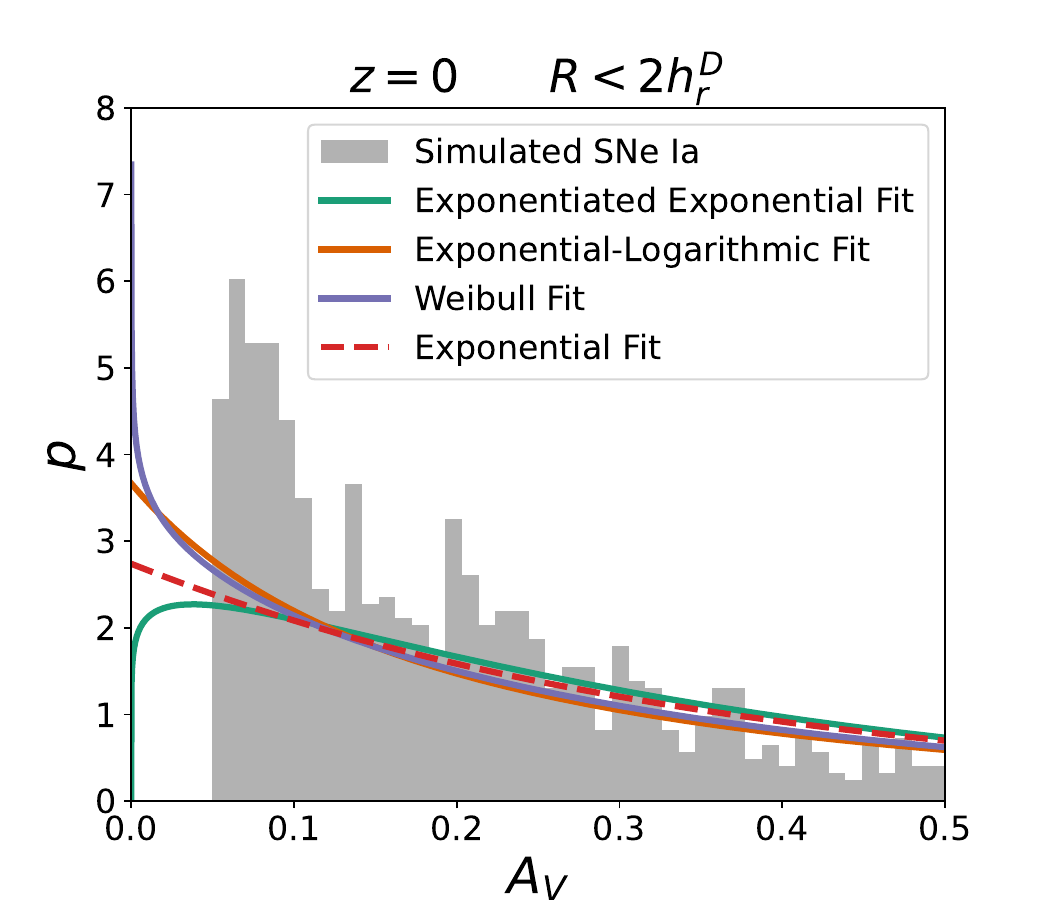}
	\caption{Same as Fig. \ref{fig:av_dist_1e7}, but considering only SNe hosted in the central region of the spiral galaxy disk.}

\label{fig:av_dist_coords}
\end{figure}

For the most part, \cite{hallgren2025} also confirm our results regarding a strongly peaked distribution not well described by an exponential. In addition, they find that, if the SNe are restricted to the central regions of the host, there is no peak at $A_V=0$, but rather a bump for some positive $A_V$. \cite{wojtak_2023} find a similar bumped extinction distribution from SN Ia cosmological fits assuming a Gamma PDF for $E(B-V)$, particularly for high-stretch SNe. As shown in Fig. \ref{fig:av_dist_coords}, we recover this same bumpy distribution if we restrict our sample to the inner regions of the spiral disk ($R<2h^{\mathrm{D}}_R$, $z=0$). We note that, while both the exponentiated exponential and the Weibull PDFs can be used to describe PDFs with $p(A_V=0)=0$, which occurs for values of $\alpha > 1$ and $\gamma > 1$, respectively, they do not result in good fits of the data for this spatially-restricted population.

\par
The inherent variation of the extinction distribution for SNe in different environments is also well documented. \cite{Hatano_1998} show that there are significant differences in the extinction distribution between SNe in the bulge and the disk within the same galaxy. Likewise, \cite{hallgren2025} report that not only do spiral, lenticular, and elliptical galaxies exhibit different $E(B-V)$ PDFs, but also that these differences persist even between spirals with different ages or bulge-to-total mass ratios. \cite{Holwerda_2014} caution that factors such as inclination and galaxy type can influence the shape of observed extinction PDFs and \cite{wojtak_2023} recover different extinction distributions for low- and high-stretch SNe.
\par
As seen in Section \ref{sec:results_I}, the exponential PDF cannot accurately capture the differences between the extinction distributions of SN populations in different galaxies and environments. One of the main reasons for this is that, due to the more restricted parameter space, host morphology and dust mass are forcibly encoded by the same parameter $\tau$. Therefore, a two-parameter description becomes necessary not only to properly reflect the extinction distribution in specific SN populations, but also to account for the differences in the extinction distributions of samples associated with different host environments.

\subsection{Impacts of the extinction distribution on SN Ia simulations, light-curve fits and cosmology}
\label{sec:impacts}
In Section \ref{sec:pantheon}, we briefly outlined some of the consequences of assuming different extinction PDFs on SN Ia intrinsic colors. In this section, we offer some comments on how this assumption can impact the results of both SN simulations and cosmology.

\par
Simulations are commonly employed as a way to study both individual events and large samples of SNe Ia. Their applications include the computation of both bias corrections for SNe Ia cosmology \citep[e.g.,][]{Kessler_2009,Kessler_2017,Popovic_2021,Vincenzi_2021} and expected SNe Ia rates \citep[e.g.,][]{Perrett_2012,Rodney_2014,wiseman_2021}. Many of these simulations assume that extinction follows either an exponential PDF \citep[e.g.,][]{Popovic_2021,wiseman_2021} or an exponential with a Gaussian component centered at $E(B-V)=0$ \citep[e.g.,][]{Rodney_2014,Vincenzi_2021}. However, according to the results presented in Section \ref{sec:results}, these assumptions do not accurately describe the true dust contributions to observed SN Ia color.
\par
In general, as stated in Section \ref{sec:intro}, intrinsic color and dust have different impacts on simulated SN magnitudes. Given two SNe with the same color $c$ at the same position, the one with the larger $E(B-V)$ will be intrinsically bluer and appear fainter \citep[e.g.,][]{Mandel_2017}. Altering the extinction PDF used in a simulation might shift the magnitudes of a given SN either above or below the detection threshold, therefore shifting the values of the simulated rates and/or biases. For example, if one of the two-parameter PDFs is used, the number of low-extinction SNe is expected to be much larger, which would make them brighter on average. As such, the simulated biases would be smaller than those computed using the standard exponential extinction PDF, while the expected observation rates would increase.

\par
Another important advantage of the two-parameter extinction PDFs is that they allow for accurate modeling of SNe in different host environments. As shown in Section \ref{sec:results_I}, the extinction distribution is strongly linked to the SN host galaxies and the exponential PDF lacks the freedom to accurately capture and describe differences in the SN environments. Using a two-parameter PDF would allow one to link different host environments to specific distribution shapes, which can improve the accuracy of any SNe simulation.

\par

The PDFs described in Tab. \ref{tab:BIC} can be used to simulate individual typical SN environments, with $\tau$ adjusted to reflect different host dust masses. The extinction for the SNe in the spiral disk is best described by the Weibull PDF, whereas SNe in the spiral bulge and the elliptical galaxy show a preference for the exponential-logarithmic PDF. However, given that the EL PDF often requires very small $\theta$ values, it can prove advantageous to use either the exponentiated exponential or Weibull PDFs instead, as they appear to be more stable. While the specific values that describe a truly universal extinction distribution cannot be inferred from our simple simulations, Figs. \ref{fig:av_dist_all_mass} and \ref{fig:av_dist_all_morph} show that its overall shape can most likely be well described by the two-parameter PDFs. Furthermore, Tabs. \ref{tab:BIC_all_mass} and \ref{tab:BIC_all_morph} offer some insight into the typical orders of magnitude that can be expected for both $\tau$ and the morphological parameters.

\par

A correct definition of the extinction distribution is also essential for both light-curve \citep[e.g.,][]{Mandel_2022} and cosmological fits \citep[e.g.,][]{Popovic_2023}. These are often based on a Bayesian framework, which requires the specification of prior PDFs for all free parameters in the model. For light-curve fitting, different extinction priors can greatly impact the values recovered for both $c_{\mathrm{int}}$ and $E(B-V)$. In the case of cosmology, as discussed before, the shift in magnitude induced by a difference in intrinsic color is not the same as the one induced by a difference in color-excess. This is the reason why the standard \cite{Tripp98} color-luminosity correction should be separated into its components. Therefore, different extinction priors will not only affect the $E(B-V)$ and $c_{\mathrm{int}}$ posteriors, but will also affect the strength of the color-luminosity correction and impact the recovered cosmological parameters \citep[e.g.,][]{wojtak_2023}.
\par
The Weibull and the exponentiated exponential PDFs have proved able to accurately fit the three SN populations studied in this paper with relatively well constrained parameters. In addition, they also have the freedom to accommodate different distribution shapes, such as a possible bump for $A_V\neq0$, even if they do not perform very well for our example of a SN population confined to the inner regions of the dust disk, as seen in Fig. \ref{fig:av_dist_coords}. For these reasons, we conclude that they offer the best descriptions of the extinction distribution.

\section{Conclusions}
\label{sec:conclusions}
In this work, we used radiative transfer simulations to explore the parameterization of dust extinction in SN Ia observations. Our main findings are as follows:

\begin{enumerate}[i]
    \item SN Ia extinction distributions cannot be properly described by an exponential PDF, even for idealized galaxy geometries. This description underestimates the abundance of low-$A_V$ SNe and overestimates the abundance of high-$A_V$ SNe.
    \item A two-parameter generalization of the exponential PDF, such as the exponentiated exponential (Eq. \ref{eq:dist_ee}), exponential-logarithmic (Eq. \ref{eq:dist_el}), Weibull (Eq. \ref{eq:dist_w}) or gamma distribution (App.~\ref{app:gamma}), offers a better description of simulated SN Ia extinction. We find that variations in host dust mass are primarily linked to variations in $\tau$, while variations in the host structural parameters and star-to-dust geometry are tied to variations in $\alpha$, $\theta$, and $\gamma$, with higher values linked to more extended dust environments or more centrally located SNe.
    \item Different SN Ia samples, originating in different host environments or within different regions of the same galaxy, are subject to different extinction distributions, which the exponential PDF cannot accurately describe. This is due to the fact that, for this PDF, the effects of host structural parameters and dust mass are described by the same parameter $\tau$. 
    \item In general, we find that the extinction distribution for SNe originating in a spiral disk is best described by a Weibull PDF, while the extinction PDF for SNe originating in a spiral bulge or an elliptical galaxy are best described by an exponential-logarithmic PDF. However, given that the exponential-logarithmic fit can struggle to properly constrain the distribution parameters, we find that the Weibull or exponentiated exponential PDFs offer the best overall descriptions of extinction in the three environments, whether for simulation or for use as a prior.
    \item Changing the extinction PDF in the SN color model affects not only the inferred color excess $E(B-V)$ but also the intrinsic color $c_{\mathrm{int}}$. Using any of the two-parameter generalized PDFs results in less extincted and intrinsically redder SNe, when compared to the standard exponential PDF.
    \item Using a two-parameter PDFs to model extinction in SN Ia simulations may have significant effects on the results. More low-extinction SNe would imply brighter observed magnitudes, which, in general, would mean a decrease of simulated SN cosmological biases and an increase of simulated SN rates. Likewise, the use of a two-parameter PDF as an extinction prior in SN Ia cosmology could potentially alter both the strength of the color-luminosity correction and the values recovered for the cosmological parameters.
\end{enumerate}

\begin{acknowledgements}
JD, AMM, and RPS acknowledge support by FCT for CENTRA through grant No. UID/PRR/00099/2025 (https://doi.org/10.54499/UID/PRR/00099/2025) and grant No. UID/00099/2025 (https://doi.org/10.54499/UID/00099/2025). JD acknowledges support by FCT under the PhD grant 2023.01333.BD, with DOI https://doi.org/10.54499/2023.01333.BD. RPS acknowledges support by FCT under the PhD grant 2024.03599.BD. This work was supported by research grants (VIL16599,VIL54489) from VILLUM FONDEN.
\end{acknowledgements}

\bibliographystyle{aa}
\bibliography{aanda.bib}

\begin{appendix}
\onecolumn
\section{Comparison with the gamma distribution}
\label{app:gamma}
In addition to the exponentiated exponential, the exponential-logarithmic, and the Weibull distributions introduced in Section \ref{sec:methods}, a common two-parameter generalization of the exponential PDF is given by the gamma distribution, whose PDF follows Eq. \ref{eq:dist_gamma} for $\gamma > 0$:

\begin{equation}
\label{eq:dist_gamma}
\centering
p_{\mathrm{Gamma}}(A_V; \tau, \gamma)=\frac{1}{\tau^{\gamma}\Gamma(\gamma)} A^{\gamma-1}_V \exp(\frac{-A_V}{\tau}),
\end{equation}
\noindent
where $\Gamma(\gamma)$ is the gamma function. This PDF reduces to the standard exponential PDF for $\gamma=1$.
\par

\begin{figure}[h]
	\centering
	\includegraphics[width=1\textwidth]{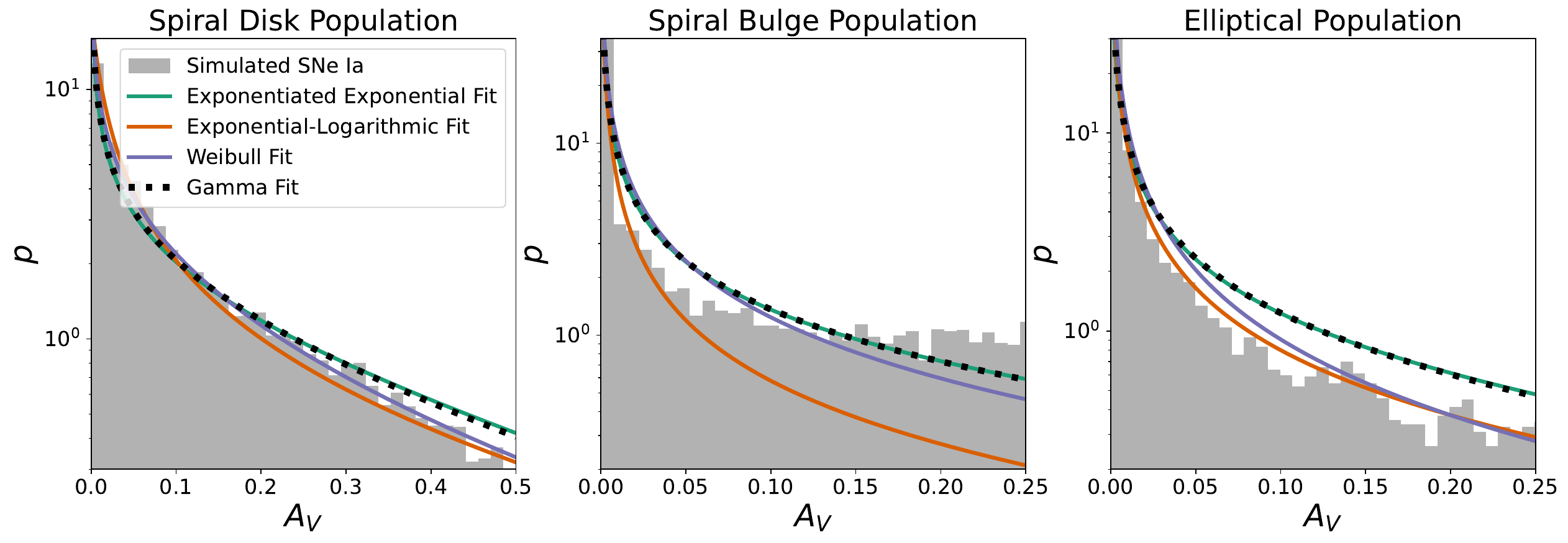}
	\caption{Same as Fig. \ref{fig:av_dist_1e7}, but including best-fit curves for the gamma distribution.}

\label{fig:av_dist_gamma}
\end{figure}

The use of this function as a parametrization for SN Ia extinction is thoroughly explored in \cite{hallgren2025}. In this section, we briefly discuss how well it can describe our simulated data and how it compares to the other parametrizations discussed above. To this effect, the simulated extinction distributions for the SN samples discussed in Section \ref{sec:results_I}, with host galaxies with fixed structural parameters and a fixed dust mass of $M_{\mathrm{D}}=10^7 M_{\odot}$, are shown in Fig. \ref{fig:av_dist_gamma}, along with best-fit curves for the gamma, EE, EL, and W PDFs. Corresponding best-fit parameters and BIC values for the gamma and EE PDFs are shown in Table \ref{tab:BIC_gamma}.

\renewcommand{\arraystretch}{1.5}
\begin{table}[h]
 \caption{Same as Tab. \ref{tab:BIC}, but with best-fit parameters and corresponding BIC values for the gamma distribution.}
    \centering
    \begin{tabular}{c c c c c c c}
    \hline\hline
         SN Environment & Model &  BIC & $\tau$ & $\alpha$ & $\gamma$ \\ \hline 
        \multirow{2}{*}{Spiral Disk} & EE  & $-1.92\times10^{4}$ & $0.411^{+0.006}_{-0.006}$ & $0.484^{+0.005}_{-0.004}$ & -\\
        & Gamma & $-1.94\times10^{4}$ & $0.491^{+0.008}_{-0.008}$ & - & $0.492^{+0.004}_{-0.004}$\\ \hline
                \multirow{2}{*}{Spiral Bulge} & EE & $1.01\times10^{8}$ & $0.720^{+0.014}_{-0.014}$ & $0.235^{+0.002}_{-0.002}$ & - \\
        & Gamma & $1.01\times10^{8}$ & $1.04^{+0.02}_{-0.02}$  & - & $0.237^{+0.002}_{-0.002}$ \\\hline
        
        \multirow{2}{*}{Elliptical Galaxy} & EE & $-9.54\times10^{4}$ & $0.470^{+0.009}_{-0.009}$ & $0.186^{+0.002}_{-0.002}$  & -  \\
        & Gamma & $-9.54\times10^{4}$ & $0.66^{+0.01}_{-0.01}$  & - & $0.189^{+0.002}_{-0.002}$  \\\hline
    \end{tabular}
    \tablefoot{Values for the exponentiated exponential PDF are transposed from Tab. \ref{tab:BIC} for comparison.}
    \label{tab:BIC_gamma}
\end{table}

\par

From both Fig. \ref{fig:av_dist_gamma} and Tab. \ref{tab:BIC_gamma}, we find that, for the three samples under study, the gamma distribution closely resembles the exponentiated exponential PDF. While the $\tau$ parameter changes substantially depending on the parameterization, $\alpha$ and $\gamma$ share very similar values. In addition, the BIC values for both PDFs are virtually identical.
\par
Fitting the gamma distribution to the simulated extinction distributions recovered for the random-host dust mass and random-host morphology samples, detailed in Sections \ref{sec:dust_mass} and \ref{sec:morph}, respectively, we find very similar results. In all instances, the gamma distribution behaves almost indistinguishably from the exponentiated exponential PDF. Likewise, the results of the Pantheon+SH0ES color fit using a gamma extinction distribution are consistent with those obtained using the EE PDF, for both $c_{\mathrm{int}}$ and $E(B-V)$.
\par
While only tentative, all these results seem to indicate that, for practical purposes, the description of extinction given by the gamma distribution is equivalent to the one offered by the exponentiated exponential PDF. We highlight, however, that the values of the parameters for both PDFs might not be exactly the same.

\FloatBarrier
\section{Mass distributions for DustPedia galaxies}
\label{app:mass_dist}

\begin{figure}[h]
	\centering
	\includegraphics[width=\textwidth]{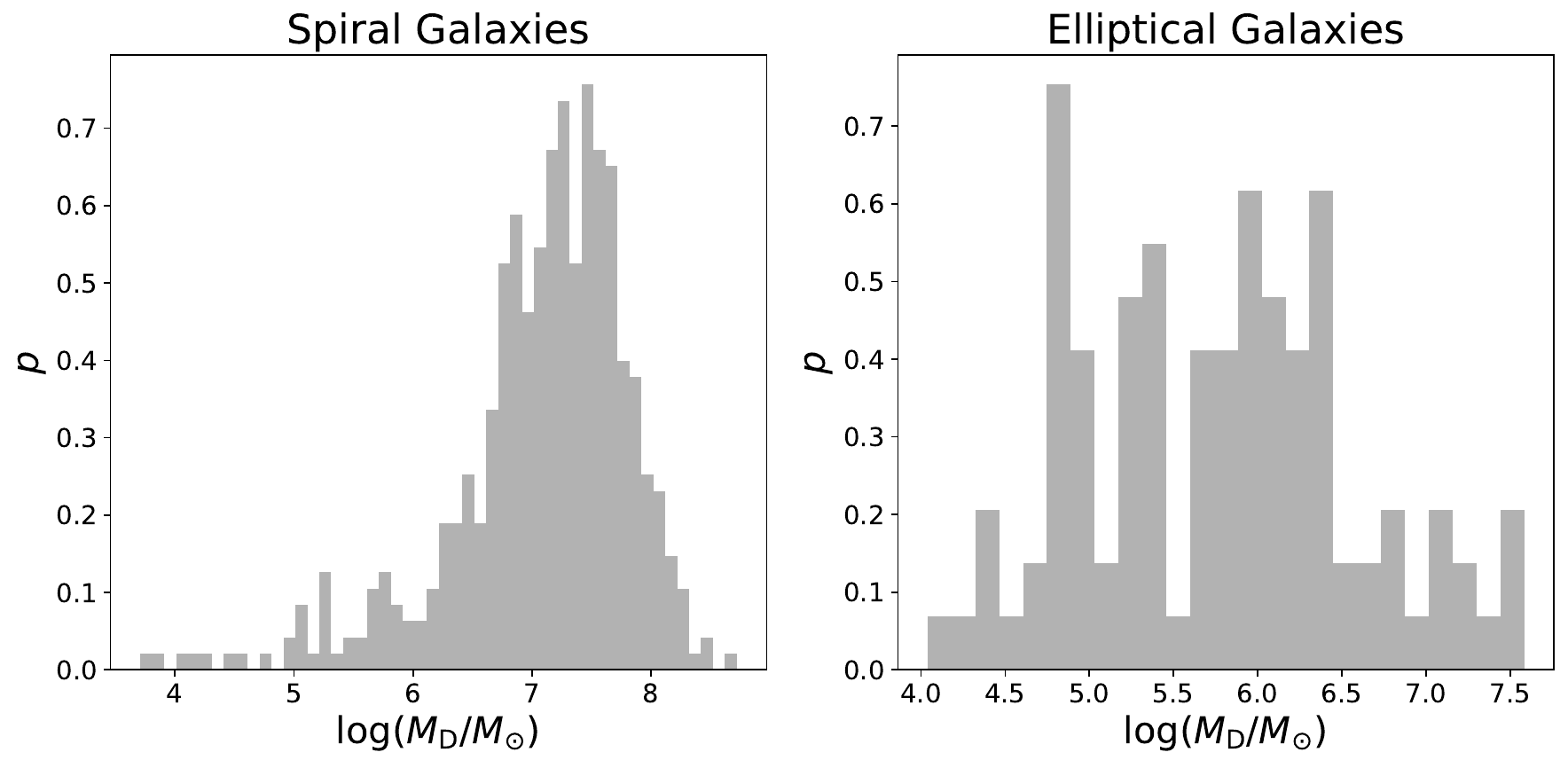}
	\caption{Dust mass distributions for spiral (left) and elliptical galaxies (right) from the DustPedia dataset.}
\label{fig:dust_mass_dist}
\end{figure}

The dust mass distributions for both spiral and elliptical galaxies from the DustPedia dataset are plotted in Fig. \ref{fig:dust_mass_dist}. For spiral galaxies, the dust mass lower bound is around $10^4 M_{\odot}$, but the majority of galaxies have $M_{\mathrm{D}}$ values in the range of $10^6 M_{\odot}$ to $10^9 M_{\odot}$. For elliptical galaxies, the dust masses range from $10^4 M_{\odot}$ to $10^8 M_{\odot}$.

\end{appendix}

\end{document}